\documentclass[paper, 12pt, letterpaper]{JHEP} 
\usepackage{amsmath}
\usepackage{amssymb}

\def\Bbb{\mathbb} \def\C{{\Bbb C}}  \def\Z{{\Bbb Z}}

    \def\tr{\operatorname{tr}}\def\Tr{\operatorname{Tr}}
\def\str{\operatorname{str}}
\def\Str{\operatorname{Str}}

 \def\tr{{\rm tr\, }} 
 \def\Tr{{\rm Tr\, }} 
   \def\deg{{\rm deg}} \def\rk{{\rm rk}}

\def\id{\protect{{1 \kern-.28em {\rm l}}}}

\newcommand{\be}{\begin{equation}} \newcommand{\ee}{\end{equation}}
\newcommand{\bea}{\begin{eqnarray}} \newcommand{\eea}{\end{eqnarray}}
\newcommand{\beann}{\begin{eqnarray*}}
  \newcommand{\eeann}{\end{eqnarray*}}
\newcommand{\bfig}{\begin{figure}} \newcommand{\efig}{\end{figure}}
\newcommand{\nn}{\nonumber}
\newcommand{\ba}{\begin{array}}\newcommand{\ea}{\end{array}}

\newtheorem{Proposition}{Proposition}[section]

\newtheorem{Theorem}{Theorem}[section]
\newtheorem{Lemma}{Lemma}[section]
\newtheorem{Corrolary}{Corrolary}[section]

\newcommand{\bp}{\begin{Proposition}}
  \newcommand{\ep}{\end{Proposition}}
\newcommand{\bt}{\begin{Theorem}} \newcommand{\et}{\end{Theorem}}
\newcommand{\bl}{\begin{Lemma}} \newcommand{\el}{\end{Lemma}}
\newcommand{\bc}{\begin{Corrolary}} \newcommand{\ec}{\end{Corrolary}}

   \def\ep{\eps}

\author{C. I.  Lazaroiu\\
  5243 Chamberlin Hall\\
  University of Wisconsin\\
  1150 University Ave\\
  Madison, Wisconsin 53706,USA\\
  calin@physics.wisc.edu
}

\title{On the boundary coupling of topological Landau-Ginzburg models}

\abstract{I propose a general form for the boundary coupling of 
  B-type topological Landau-Ginzburg models.  In particular, I show that the
  relevant background in the open string sector is a (generally non-Abelian)
  superconnection of type $(0,1)$ living in a complex superbundle defined on
  the target space, which I allow to be a non-compact Calabi-Yau manifold.
  This extends and clarifies previous proposals. Generalizing an argument due
  to Witten, I show that BRST invariance of the partition function on the
  worldsheet amounts to the condition that the $(0,\leq 2)$ part of the
  superconnection's curvature equals a constant endomorphism plus the
  Landau-Ginzburg potential times the identity section of the underlying
  superbundle. This provides the target space equations of motion for the open
  topological model.}

\preprint{MAD-TH-03-6}

\begin{document}

\tableofcontents

\pagebreak

\vskip .6in

\section{Introduction}
\label{intro}

It was recently pointed out \cite{Kontsevich, Kap1, Orlov, Lerche, Kap2, Kap3}
that Landau-Ginzburg models contain many more topological B-type branes
than previously believed. This is potentially important since 
it seems to allow for a simple realization of the framework of \cite{Konts,
  Moore_Segal, Moore, CIL1, CIL2, CIL3, CIL4, CIL5, Douglas, Aspinwall, Sharpe} in such theories. 

The physics argument of \cite{ Kap1, Lerche, Kap2, Kap3} involves the
construction of a boundary action for the untwisted (i.e. $N=2$)  
model, whose B-type supersymmetry variation is shown to
cancel the boundary term produced by the variation of the bulk
action.  The object described by this coupling can be understood as a
composite of two elementary branes obtained by condensing a tachyon. 
More precisely, the Landau-Ginzburg model reduces to an $N=2$ sigma model when one
turns off the Landau Ginzburg potential $W$. Starting with two top rank 
B-branes of the sigma model and turning on $W$ forces the condensation of a
tachyon. This leads to a D-brane composite modeled by the boundary action 
considered in \cite{ Kap1, Lerche, Kap2, Kap3}. Thus the basic branes
of Landau-Ginzburg models are composites of elementary branes of the
associated {\em sigma model}, with tachyon condensation driven by the potential $W$! 
This point of view leads to novel insight into the D-brane dynamics of such
theories \cite{us}. 

Unfortunately, the construction of the boundary coupling has been carried out only under
simplifying assumptions, which must be removed in order to make further
progress. Among the limitations of previous analyses 
are the condition that the target space of the model is 
an affine space $\C^n$, the fact that the boundary action is known only when 
both underlying sigma model branes have multiplicity one
(i.e. they carry {\em Abelian} gauge connections on their worldvolumes), and the fact
that gauge degrees of freedom on these branes are not included. 
Finally, the boundary action is constructed in the untwisted,
$N=2$ model, which makes it difficult to apply off-shell techniques or 
perform a systematic study of localization.  

In the present note, I remove each of these limitations by proposing a general
boundary coupling for the topological Landau-Ginzburg model with an arbitrary
noncompact Calabi-Yau target $X$. As it turns out,
the relevant open string background is a $(0,1)$ superconnection in 
a complex superbundle $E$ over $X$. This matches expectations from string
field theory and known results from the B-twisted sigma model
\cite{Witten_CS,
  CIL3, Diaconescu}. By extending an argument of \cite{Witten_CS} to the case
of superconnections, I show that BRST invariance of the worldsheet partition
function is restored by this coupling provided that the $(0,\leq 2)$ part of the superconnection's
curvature equals a constant endomorphism plus $W$ times the identity
endomorphism of $E$. This provides the target space equations of motion of the
topological open string background. The construction of the present paper is
carried out directly in the twisted model. This leads to a few
simplifications, and in particular allows us to avoid the introduction of
certain complex conjugate terms.

\paragraph{Observation} Strictly speaking, 
our discussion will be classical on the worldsheet, so we could allow $X$ 
to be only K\"{a}hler. However anomaly cancellation requires that $X$ is
Calabi-Yau in order for the theory to make sense at the quantum level 
\cite{Witten_mirror}.

\section{The bulk action and the topological Warner term}
\label{bulk}

The general formulation of closed B-type topological Landau-Ginzburg models was given a
while ago in \cite{Labastida} by extending the work of
\cite{Vafa_LG}. The target space of such models is a Calabi-Yau manifold $X$, 
and the Landau-Ginzburg potential is a holomorphic function $W\in H^0({\cal O}_X)$. 
Since any holomorphic function on a compact complex manifold is constant, we
must assume that $X$ is non-compact in order to obtain interesting models. 
As for the B-twisted sigma model \cite{Witten_mirror}, the 
Grassmann even worldsheet fields of the on-shell formulation are given by
the components $\phi^i$, $\phi^{\bar i}$ of the map $\phi:\Sigma \rightarrow
X$, while the G-odd fields are sections $\eta$, $\theta$ 
and $\rho$ of the bundles $\phi^*({\bar  T}X)$, 
$\phi^*(T^*X)$  and $\phi^*(TX)\otimes {\cal T}^*\Sigma$ over the worldsheet
$\Sigma$. Here ${\cal T}^*\Sigma$ is the complexified cotangent bundle to $\Sigma$, while
$TX$ and ${\bar T}X$ are the holomorphic and antiholomorphic components of the
complexified tangent bundle ${\cal T}X$ to $X$.

To write the bulk topological action, we introduce new fields $\chi, {\bar
  \chi} \in \Gamma(\Sigma, \phi^*({\bar T}X))$ by the relations:
\bea
\eta^{\bar i}&=&\chi^{\bar i}+{\bar \chi}^{\bar i}\\
\theta_i&=&G_{i{\bar j}}(\chi^{\bar j}-{\bar \chi}^{\bar j})~~.
\eea
We will also use the quantity $\theta^{\bar i}=G^{{\bar i}j}\theta_j$.

As explained in \cite{Labastida}, it is convenient to use an off-shell
realization of the BRST symmetry. For
this, we introduce an auxiliary G-even field ${\tilde F}$ which transforms as a section of
$\phi^*({\cal T}X)$. Then the BRST transformations are:
\bea
\label{BRST}
\delta \phi^i=0~~&,&~~\delta \phi^{\bar i}=\chi^{\bar i}+{\bar \chi}^{\bar
  i}=\eta^{\bar i}\nn\\
\delta \chi^{\bar i}={\tilde F}^{\bar i}-\Gamma^{\bar
  i}_{{\bar j}{\bar k}}{\bar \chi}^{\bar j}\chi^{\bar k}~~&,&~~\delta {\bar \chi}^{\bar
  i}=-{\tilde F}^{\bar i}+\Gamma^{\bar i}_{{\bar j}{\bar
  k}}{\bar \chi}^{\bar j}\chi^{\bar k}~~\nn\\
\delta \rho^i_\alpha&=&2\partial_\alpha \phi^i~~\\
\delta {\tilde F}^i=i\varepsilon^{\alpha\beta} \left[D_\alpha
  \rho^i_\beta+\frac{1}{4}R^i_{j{\bar l}k}(\chi^{\bar l}+{\bar \chi}^{\bar l})
\rho^j_\alpha\rho^k_\beta\right]~~&,&~~\delta
{\tilde F}^{\bar i}=\Gamma^{\bar i}_{{\bar j}{\bar k}}{\tilde F}^{\bar j}(\chi^{\bar k}+{\bar \chi}^{\bar k})~~.\nn
\eea
Notice that the off-shell BRST transformations are independent of
  $W$. Moreover, the transformations of $\phi$, $\eta$ and $\rho$ 
do not involve the auxiliary fields. In particular, we have  $\delta \eta^{\bar i}=0$.
These observations will be used in Section 4. 

Let us choose a Riemannian metric $g$ on the  worldsheet. Then the bulk action
of \cite{Labastida} is:
\be
\label{Sbulk_decomp}
S_{bulk}=S_B+S_W
\ee
where:
\bea
\label{S_B}
\!\!\!\!\!S_B&=&\int_\Sigma {d^2\sigma \sqrt{g}}{\Big[
G_{i{\bar j}}\left(g^{\alpha\beta} \partial_\alpha \phi^i \partial_\beta
  \phi^{\bar j} -i\varepsilon^{\alpha\beta}\partial_\alpha
  \phi^i\partial_\beta\phi^{\bar j}-\frac{1}{2}g^{\alpha\beta}\rho_\alpha^i
  D_\beta \eta^{\bar j}-\frac{i}{2}\varepsilon^{\alpha\beta} \rho_\alpha^i
  D_\beta \theta^{\bar j}-{\tilde F}^i{\tilde F}^{\bar j}\right)}\nn\\
\!\!\!\!&+&\frac{i}{4}\varepsilon^{\alpha\beta}R_{i{\bar l}k{\bar
      j}}\rho^i_\alpha{\bar \chi}^{\bar l}\rho^k_\beta\chi^{\bar j}\Big]
\eea
is the action of the B-twisted sigma model in the form used in that reference 
and $S_W=S_0+S_1$ is the potential-dependent term, with:
\bea
S_0&=&\frac{i}{2}
\int_{\Sigma}{d^2\sigma \sqrt{g} \left[D_{\bar i} \partial_{\bar j}{\bar W} \chi^{\bar
      i}{\bar \chi}^{\bar j}-(\partial_{\bar i}{\bar W}){\tilde F}^{\bar i}\right]}\\
S_1&=&-\frac{i}{2}\int_{\Sigma}{d^2\sigma \sqrt{g} \left[(\partial_i W){\tilde F}^i +\frac{i}{4} 
\varepsilon^{\alpha\beta}D_i\partial_jW \rho_\alpha^i\rho_\beta^j\right]}~~.
\eea
Here $\varepsilon^{\alpha\beta}=\frac{\epsilon^{\alpha\beta}}{\sqrt{g}}$ is
the Levi-Civita tensor and  $\epsilon^{\alpha\beta}$ the associated
density. We have rescaled the Landau-Ginzburg potential $W$ by a factor of
$\frac{i}{2}$ with respect to the conventions of \cite{Labastida}. The
conventions for the target space Riemann tensor and 
covariantized worldsheet derivative $D_\alpha$ are unchanged. 
In $S_W$, we separated the term depending on
$W$ from that depending on its complex conjugate ${\bar W}$. 

It was noticed in \cite{Labastida} that the topological sigma model action in
the form (\ref{S_B}) is BRST exact on closed Riemann surfaces.
Since in this paper we shall allow $\Sigma$ to have a nonempty boundary, we
must be careful with total derivative terms. Extending the computation of
\cite{Labastida} to this case, one finds:
\be
\label{exactS_B}
S_B+s=\delta V_B
\ee
where:
\be
V_B:=\int_{\Sigma}{d^2\sigma
  \sqrt{g} G_{i{\bar j}}\left(\frac{1}{2}g^{\alpha\beta}\rho^i_\alpha\partial_\beta \phi^{\bar
      j}-\frac{i}{2}\varepsilon^{\alpha\beta}\rho^i_\alpha
    \partial_\beta\phi^{\bar j}-{\tilde F}^i\chi^{\bar j}\right)}
\ee
and:
\be
s:=i\int_{\Sigma}{d^2\sigma \sqrt{g}\varepsilon^{\alpha\beta}\partial_\alpha(G_{{\bar
      i}j}\chi^{\bar i}\rho^j_\beta)}=i\int_\Sigma{d(G_{{\bar
        i}j}\chi^{\bar i}\rho^j)}
\ee
is a total derivative term. Since such a term does not change physics on
closed Riemann surfaces, we are free to redefine the bulk topological sigma-model
action by adding it to $S_B$:
\be
{\tilde S}_B:=S_B+s=\delta V_B~~.
\ee
Accordingly, we shall work with the modified bulk Landau-Ginzburg action:
\be
{\tilde S}_{bulk}=S_{bulk}+s={\tilde S}_B+S_0+S_1~~.
\ee
It is easy to check that the term $S_0$ is also BRST exact:
\be
\label{exactS_0}
S_0=\delta V_0
\ee
where:
\be
V_0=-\frac{1}{2}\int_{\Sigma}{d^2\sigma \sqrt{g}\theta^{\bar i} \partial_{\bar i}{\bar W}}~~.
\ee
Equations (\ref{exactS_B}) and (\ref{exactS_0}) are local, i.e. they hold for the 
associated Lagrange densities without requiring integration by parts. Thus
both of these relations can be applied to bordered Riemann surfaces. 

It is now easy to show that the BRST variation of ${\tilde S}_{bulk}$ produces a boundary term:
\be
\label{deltaSbulk}
\delta {\tilde S}_{bulk} =\delta S_1=\frac{1}{2} \int_{\partial \Sigma}{\rho^i\partial_i
  W}~~.
\ee
The presence of a non-zero right hand side in (\ref{deltaSbulk}) is known as the Warner
  problem \cite{Warner}. As we shall see below, this term induces a
  deformation of the target space equations of motion in the open string sector.

\section{The boundary coupling}
\label{boundary_coupling}

Since B-type Landau-Ginzburg models are obtained from topological sigma models with
non-compact targets by adding a potential term, it is natural to look for
their open string backgrounds as deformations of the backgrounds allowed in
the sigma model. It is well-known \cite{Hori} that B-twisted Landau-Ginzburg
models do not admit interesting elementary branes of top dimension. Therefore,
one should look for D-brane composites obtained by condensing fields between
brane-antibrane pairs of the {\em sigma model}.  As explained in \cite{CIL2, CIL3,
Diaconescu}, topological D-brane composites of the B-twisted\footnote{A
similar result exists for the A-twisted sigma model. In that case, graded
superconnections of total degree one in the sense of \cite{Bismut_Lott} are
the target space fields in the diagonal sector of Fukaya's category
\cite{Vafa_CS, CIL6,CIL7,CIL8,CIL9,CIL10, Fukaya}.}  sigma model are described by a
Dolbeault version of the 'graded superconnections of total degree one' defined
in \cite{Bismut_Lott}. Such objects are generalizations of the more familiar
superconnections of type $(0,1)$.  Graded rather than standard
superconnections appear for the sigma model because topological D-branes are
$\Z$-graded in that context \cite{Konts,Seidel, CIL3, Douglas, Diaconescu}.  
As explained in \cite{Kap1, Kap2}, turning on the Landau-Ginzburg
potential will generally break the integral grading to a $\Z_2$ group. Thus
one expects that B-type Landau-Ginzburg branes are described by $(0,1)$
superconnections, a variant of the objects considered in \cite{Quillen}.  As
we explain below, this expectation is indeed realized.

To be precise, we consider a complex superbundle $E=E_+\oplus E_-$ over $X$, and a
superconnection ${\cal B}$ on $E$. We let $r_\pm:=\rk E_\pm$. Remember that
the bundle of endomorphisms $End(E)$ 
has a natural $\Z_2$ grading, whose even and odd components are
given by:
\bea
End_+(E)&:=&~~~~~End(E_+)\oplus End(E_-)\\
End_-(E)&:=& Hom(E_+,E_-)\oplus Hom(E_-,E_+)~~.
\eea
Then ${\cal B}$  can be viewed 
as a section of the bundle $[{\cal T}^*X\otimes End_+(E)]\oplus End_-(E)$. 
In a local frame of $E$ compatible with the grading, this is simply 
a matrix:
\be
{\cal B}=\left[\ba{cc} A^{(+)}& F\\G & A^{(-)}\ea\right]
\ee
whose diagonal entries $A^{(\pm)}$ are connection one-forms on $E_\pm$, while 
$F,G$ are elements of  $Hom(E_-,E_+)$ and $Hom(E_+,E_-)$.  As for the B-model, we 
require\footnote{One can remove this condition, but in that case the $(1,0)$
part of $A^{(\pm)}$ would be non-dynamical.} 
that the superconnection has type $(0,1)$, i.e. the one-forms $A^{(\pm)}$  belong to
$\Omega^{(0,1)}(End(E_\pm))$. The morphism $F$ should not be
confused with the curvature form used below.

When endowed with the ordinary composition of morphisms, the space of sections
$\Gamma(End(E))$ becomes an associative superalgebra. The space
${\cal H}_0=\Omega^{(0,*)}(End(E))$ also carries an associative superalgebra structure,
which is induced from $(\Omega^{(0,*)}(X),\wedge)$ and 
$(\Gamma(End(E)),\circ)$ via the tensor product decomposition:
\be
\Omega^{(0,*)}(End(E))=\Omega^{(0,*)}(X)\otimes_{\Omega^{(0,0)}(X)} \Gamma(End(E))~~.
\ee
Note that we take form components to sit on the left. For decomposable
elements $u=\omega\otimes f$ and $v=\eta\otimes g$, with
homogeneous\footnote{Homogeneity of $\omega$ and $\eta$ means that both are
  differential forms of given rank (rather than sums of forms of different ranks).}  
$\omega,\eta$ and $f,g$, the product $\bullet$ on ${\cal H}_0$ takes the form:
\be
u\bullet v=(-1)^{ \deg f~\rk \eta}(\omega\wedge \eta)\otimes (f\circ g)~~,
\ee
where $\deg$ denotes the $\Z_2$-valued degree in the superalgebra $End(E)$:
\be
\deg(f)=0~~{\rm~if~}f\in End_+(E)~~,~~\deg(f)=1~~{\rm~if~}f\in End_-(E)~~.
\ee
The total degree on ${\cal H}_0$ is given by:
\be
|\omega\otimes f|=\rk \omega +\deg f~~(mod~2)~~.
\ee
We also recall the supertrace on $End(E)$:
\be
\label{str}
\str(f)=\tr f_{++}-\tr f_{--}~~.
\ee
Here $f=\left[\ba{cc}f_{++}&f_{-+}\\f_{+-}&f_{--}\ea\right]$ is an
endomorphism of $E$ with components $f_{\alpha\beta}\in Hom(E_\alpha,E_\beta)$
for $\alpha,\beta=+,-$. The supertrace has the property:
\be
\label{str_cyc}
\str(f\circ g)=(-1)^{\deg f \deg g}\str(g\circ f)~~
\ee
for homogeneous elements $f,g$. 

Notice that ${\cal B}$ is an odd element of ${\cal H}_o$. 
Thus the twisted Dolbeault operator:
\be
\label{sc}
{\bar {\cal D}}={\overline \partial} +{\cal B}=\left[\ba{cc} 
{\bar \partial}+A^{(+)}& F\\G &{\bar \partial}+A^{(-)}\ea\right]
\ee
induces an odd derivation ${\bar \partial}+[{\cal B},\cdot]_\bullet$ 
of the superalgebra $({\cal H}_o, \bullet)$. Here 
$[u,v]_\bullet:=u\bullet v-(-1)^{|u||v|}v\bullet u$ is the supercommutator. 

The $(0,\leq 2)$ part of the superconnection's curvature has the form:
\be
\label{curvature}
{\cal F}^{(0,\leq 2)}={\bar {\cal D}}^2={\overline \partial}
{\cal B}+\frac{1}{2}[{\cal B}, {\cal B}]_\bullet={\overline \partial}
{\cal B}+{\cal B}\bullet {\cal B}=
\left[\ba{cc} F^{(+)}_{(0,2)}+FG& {\bar \nabla} F\\{\bar \nabla} G &F^{(-)}_{(0,2)}+GF\ea\right]
\ee
where $F^{(\pm)}_{(0,2)}$ are the $(0,2)$ pieces of the curvature forms
$F^{(\pm)}$ of $A^{(\pm)}$ and:
\bea
{\bar \nabla}F&=&{\bar \partial}F+
A^{(+)}\bullet F+F\bullet A^{(-)}={\bar \partial}F+A^{(+)}\circ F-F\circ A^{(-)}~~\nn\\
{\bar \nabla}G&=&{\bar \partial}G+A^{(-)}\bullet G+G\bullet A^{(+)}
={\bar \partial} G+A^{(-)}\circ G-G\circ A^{(+)}~~.
\eea

It is convenient to introduce the notations:
\be
\label{AD}
A:=A^{(+)}\oplus A^{(-)}=\left[\ba{cc} A^{(+)}& 0\\0 & A^{(-)}\ea\right]~~,~~
D:=\left[\ba{cc} 0& F\\G & 0\ea\right]
\ee
for the diagonal and off-diagonal parts of ${\cal B}$. Then $A$ is an
ordinary connection one-form on $E$ (which is compatible with the grading), while
$D$ is an odd endomorphism of $E$. We have ${\cal B}=A+D$ and:
\be
\label{02curvature}
{\cal F}^{(0,\leq 2)}=F^{(0,2)}+{\bar \nabla}_A D +D^2~~.
\ee
Here $F^{(0,2)}=F^{(+)}_{(0,2)}+F^{(-)}_{(0,2)}$ is the $(0,2)$ part of the curvature of $A$ and 
${\bar \nabla}_A={\bar \partial}+[A,\cdot]_\bullet $ is the Dolbeault operator twisted by $A$.  
Notice that $[A,D]_\bullet=dx^{\bar i}[A_{\bar i},D]$, 
where $[\cdot,\cdot]$ denotes the usual commutator. 

To couple the model to such backgrounds, we shall extend the approach used in 
\cite{Witten_CS} for coupling ordinary $(0,1)$ connections to the B-twisted
sigma model  (for comparison, that construction is reviewed in Appendix
\ref{Bmodel}).  Namely, we define the partition function on a bordered Riemann 
surface $\Sigma$ by the formula:
\be
\label{Z}
Z:=\int{{\cal D}[\phi]{\cal D}[{\tilde F}]{\cal D}[\theta]{\cal D}[\rho]{\cal D}[\eta]
  e^{-{\tilde S}_{bulk}} {\cal U}_1\dots {\cal U}_h}
\ee
where $h$ is the number of holes and the factors ${\cal U}_\alpha$ have the form:
\be
\label{calU}
{\cal U}_\alpha:=\Str Pe^{-\oint_{C_\alpha}{d\tau_\alpha M}}~~.
\ee
Here $C_\alpha$ is the boundary circle associated with the hole labeled $\alpha$,
while $\Str$ is the supertrace on $GL(r_+|r_-)$. The symbol $d\tau_\alpha$ 
stands for the length element along $C_\alpha$ induced by the metric on the
interior of $\Sigma$. The quantity $M$ is given by:
\be
\label{M}
M=\left[\ba{cc} 
{\hat {\cal A}}^{(+)}+FG &
\frac{1}{2}\rho^i_0 \nabla_i F
\\
\frac{1}{2} \rho^i_0 \nabla_i G &
{\hat {\cal A}}^{(-)} +GF 
\ea\right]~~,
\ee
where $\rho_0^i d\tau_\alpha$ is the pull-back of $\rho^i$ to $C_\alpha$ and:
\be
\label{calApm} 
{\hat {\cal A}}^{(\pm)}:=A_{\bar i}^{(\pm)}{\dot \phi}^{\bar i} +\frac{1}{2}\eta^{\bar i}  
F^{(\pm)}_{{\bar i} j}\rho_0^j
\ee
are connections on the bundles ${\cal E}_\pm$ obtained by pulling back
$E_\pm$ to the boundary of $\Sigma$. The dot in (\ref{calApm}) stands for the
derivative $\frac{d}{d\tau_\alpha}$. Notice that $\nabla_i F=\partial_i F$ and
$\nabla_i G=\partial_i G$ since $A$ is a $(0,1)$-connection.

To insure BRST invariance of (\ref{Z}), we must choose the background superconnection
${\cal B}$ such that:
\be
\delta {\cal U}_\alpha=\frac{1}{2}{\cal U}_\alpha \int_{C_\alpha}{\rho^i\partial_i W}~~. 
\ee
Then the variation of the product $\prod_{\alpha}{{\cal U}_\alpha}$ 
compensates the Warner contribution 
(\ref{deltaSbulk}) induced from the bulk.  In the next section we show by direct
computation that:
\be
\label{Qhol}
\delta {\cal U}_\alpha=-\Str \left[~I_\alpha(\delta M)
  Pe^{-\oint_{C_\alpha}{d\tau_\alpha M}}\right]
\ee
where:
\bea
&&\!\!\!\!\!\!\!I_\alpha(\delta M)=\nn\\
&&\!\!\!\!\!\!\!\!\oint_{C_\alpha}{d\tau_\alpha 
  U_\alpha^{-1}\left[F_{{\bar i}{\bar j}}\eta^{\bar i}{\dot \phi}^{\bar
      j}-\frac{1}{4} \partial_k F_{{\bar i}{\bar j}}\eta^{\bar i}\eta^{\bar
      j}\rho_0^k+\eta^{\bar i}\nabla_{\bar i} (D^2)-{\dot \phi}^{\bar
      i}\nabla_{\bar
      i}D-\frac{1}{2}\rho^i_0\partial_i(D^2)+\frac{1}{2}\eta^{\bar
      i}\rho^j_0\partial_j\nabla_{\bar i}D\right]U_\alpha}~,~~~~\nn
\eea
with $U_\alpha(\tau_\alpha)\in GL(r_+|r_-)$ a certain invertible operator playing
the role of `parallel transport' defined by $M$ along $C_\alpha$. Here
  $F_{{\bar i}{\bar j}}$ etc. are the $(0,2)$ components of 
the curvature of the direct sum connection $A$ introduced in (\ref{AD}).
Hence the BRST invariance conditions are:
\bea
&&F_{{\bar i}{\bar j}}=0\\
&&\nabla_{\bar i} D=0\\
&&\partial_i(D^2)=\partial_iW
\eea
The first relation says that $A$ is integrable, so it defines a complex
structure on the bundle $E$. The second condition means that $D\in End(E)$ is
holomorphic with respect to this complex structure. Finally, the last equation 
requires $D^2=c+W {\rm id}_E$ where $c$ is a constant endomorphism of $E$. Comparing with
(\ref{02curvature}), we see that these conditions are equivalent with:
\bea
\label{eom}
{\cal F}^{(0,\leq 2)}=c+W{\rm id}_E\Longleftrightarrow {\cal {\bar D}}^2=c+W{\rm id}_E~~.
\eea
This is the target space equation of motion for our open string background.

In the limit $W=0$ and with the choice $c=0$, relation (\ref{eom}) reduces to the condition ${\cal
  F}^{(0,\leq 2)}=0\Leftrightarrow {\bar {\cal D}}^2=0$, which is the target space
  equation of motion when coupling a topological brane-antibrane pair to the B-twisted
  sigma model. More precisely, this is a $\Z_2$ reduction of the full
  equations of motion for that model, which involve a graded superconnection
  \cite{CIL3, Diaconescu} due to the
  $\Z$-valued nature of the topological D-brane grade in that case.

Condition (\ref{eom}) generalizes particular cases established in \cite{Kap1,
 Lerche, Kap2}. Our proof is general and
 in particular works in the non-Abelian case
\footnote{The construction of \cite{Kap1, Lerche, Kap2} was carried out
 for the Abelian case $r_+=r_-=1$ and with the supplementary
 assumption that the target space is $\C^n$, in which case the connections
 $A^{(\pm)}$ can be gauged away. One also assumed that the target space metric
 is flat.}. Working
 directly with the twisted model allows us to avoid certain conjugate terms
 considered in \cite{Kap2}. Notice that we do not construct
 a boundary action. 
 From the perspective of the present paper, the approach of \cite{Kap1,
 Lerche, Kap2} is recovered when restricting to the case $r_+=r_-=1$, since in that
 situation one can replace our path ordered exponentials by path integrals over
 pairs of new fermionic fields living on the connected components of the boundary. Such a
 representation seems complicated for $r_+$ or $r_-$ greater than one.
 While interesting in its own right, it is not necessary for our purpose. 

\paragraph{Observation} As in \cite{Witten_CS}, our construction does not
 require a modification of the BRST operator. Since the boundary coupling 
introduced in (\ref{Z}) involves only the bulk worldsheet fields, its BRST
variation is computed by using the bulk generator (\ref{BRST}).

\section{BRST variation of the superholonomy factors}

In this section we prove the crucial relation (\ref{Qhol}). For this, let us
focus a given boundary component $C$, whose proper length coordinate we shall
denote by $\tau$. To specify such a coordinate, we must chose an origin on the
circle $C$. While various intermediate steps in our computation will depend
on this choice, the final result (\ref{Qhol}) is not sensitive to it.  We let
${\cal U}$ be the superholonomy factor associated to $C$, defined as in (\ref{calU}).

\subsection{Preparations}

Before proceeding with the computation, we make a few conceptual
remarks. First, notice that the object $M(\tau)$ of equation (\ref{M}) 
involves the Grassmann odd fields $\rho^i_0$.Technically, the definition of
$M$ involves a few steps. Fist, we pick a trivialization ${\cal E}=C\times V$ of the 
pulled-back superbundle over the circle $C$. Here $V=V_+\oplus V_-$ is a super-vector space 
isomorphic with the fiber of $E$. Then $M$ given in (\ref{M}) can be viewed as an even element of
the associative superalgebra ${\cal K}_e:={\cal F}\otimes End(V)$, where ${\cal F}$ is
the supercommutative algebra of superfunctions on the circle $C$. The product in
${\cal K}_e$ has the following form on decomposable elements $\phi=\alpha\otimes f$
and $\gamma=\beta\otimes g$:
\be
(\alpha\otimes f)(\beta\otimes g)=(-1)^{\deg \beta \deg f}(\alpha
\beta)\otimes (f\circ g)~~.
\ee
The total $\Z_2$-valued degree is given by $\deg(\alpha\otimes
f)=\deg\alpha+\deg f$. 

The supertrace (\ref{str}) on $End(E)$ induces the following ${\cal F}$-valued
trace on ${\cal K}_e$:
\be
\label{Str}
\Str(\alpha\otimes f)=\alpha \str(f)~~,
\ee
where $\str(f)$ is of course a complex number, while $\alpha$ is a
superfunction on the circle. $\Str$ is the supertrace appearing in equation (\ref{calU}). 
Using relation (\ref{str_cyc}), one easily checks the cyclicity property:
\be
\Str(\phi\gamma)=(-1)^{\deg \phi \deg \gamma}\Str(\gamma\phi)~~.
\ee

Viewing superfunctions on the circle as valued in a Grassmann
algebra ${\cal G}$, the elements of ${\cal K}_e$ are valued in the associative
superalgebra $K_e:={\cal G}\otimes End(V)$.
Then the supergroup $GL(r_+|r_-)$ can be viewed as the group of even and
invertible elements of $K_e$ (i.e. the group of
units of the ordinary associative subalgebra obtained by restricting to 
the even part of $K_e$). Even elements of $K_e$ have the form 
$\phi=\left[\ba{cc}\phi_{++}&\phi_{-+}\\\phi_{+-}&\phi_{--}\ea\right]$, where 
$\phi_{++}$ and $\phi_{--}$ are Grassmann even while $\phi_{+-}$ and
$\phi_{-+}$ are Grassmann odd. Restricting $\Str$ to invertible even elements 
recovers the supertrace on $GL(r_+|r_-)$.

Let us next give the precise description of the holonomy operator appearing under
the supertrace in (\ref{calU}).  For this, let $\tau\geq \tau_0$ 
and define even elements $U(\tau,\tau_0)$  of $K_e$ by the formula
\footnote{Technically, we must specify a norm
  with respect to which the defining series of $U$ is absolutely
  convergent. This can be achieved by using a Banach Grassmann algebra ${\cal G}$ to model
  the boundary fields. Then $K_e$ becomes a Banach
  algebra and absolute convergence of the series in
  (\ref{U1})  follows from continuity of $M$ along
  the compact $C$, which implies good bounds for the
  multiple integrals.}:
\be
\label{U1}
U(\tau, \tau_0)=Pe^{-\int_{\tau_0}^{\tau}{ds M(s)}}=
\sum_{n\geq 0}{(-1)^n\int_{\tau_0}^\tau{ds_1}
\int_{\tau_0}^{s_1}{ds_2}\dots
\int_{\tau_0}^{s_{n-1}}{ds_n}{M(s_1)M(s_2)\dots M(s_n)}}~~.
\ee
For $\tau_0=\tau$ we have $U(\tau_0,\tau_0)=1$. 

Using (\ref{U1}), its is easy to check the relation:
\be
\frac{\partial}{\partial \tau} U=-M(\tau)U(\tau,\tau_0)~~
\ee
as well as invertibility of $U$.
Since $M(\tau)$ is periodic
with period given by the length $l$ of $C$, we have:
\be
U(\tau+l,\tau_0+l)=U(\tau,\tau_0)~~.
\ee 
Finally, one can use definition (\ref{U1}) to check the
composition rule:
\be
U(\tau_2,\tau_1)U(\tau_1,\tau_0)=U(\tau_2,\tau_0)~~.
\ee
With these preparations, consider the `holonomy operator':
\be
H(\tau):=U(\tau+l,\tau)~~.
\ee
Then the factors ${\cal U}$ in (\ref{calU}) are defined by:
\be
{\cal U}=\Str H(\tau)~~.
\ee
To check that this is independent of $\tau$, notice that:
\be
H(\tau)=U(\tau,0)H(0)U(\tau,0)^{-1}
\ee
and use the fact that $U$ is Grassmann even.

We have to clarify one final point. When computing the BRST variation of
(\ref{calU}), we will need to move the bulk BRST operator over the
supertrace. Since the BRST action was originally given only for
superfunctions on the worldsheet, this requires that we define an 
extension $\delta_e$ of $\delta$ to the associative superalgebra ${\cal K}_e={\cal
  F}\otimes End(V)$. We shall take this extension to be given 
in the obvious manner, namely:
\be
\delta_e:=\delta \otimes {\rm id}_{End(V)}~~.
\ee
With this definition, $\delta_e$ squares to zero and 
is an odd derivation of the superalgebra ${\cal K}_e$. Moreover, we find:
\be
\delta \Str(\alpha\otimes f)=\delta(\alpha \str(f))=(\delta \alpha)
\str(f)=\Str(\delta \alpha \otimes f)=\Str(\delta_e(\alpha\otimes f))~~,
\ee
where we recall that $\str(f)$ is a just a complex number.
This gives the desired relation:
\be
\delta \Str(\phi)=\Str(\delta_e \phi)~~{\rm for}~~\phi \in {\cal K}_e~~.
\ee
This construction might seem too trivial to mention, 
but it has one important consequence. Because we shall compute the BRST
variation of ${\cal U}$ by working in the algebra ${\cal
  K}_e$, we must treat $D$ as an {\em odd} element (indeed its $\Z_2$-valued degree 
in this algebra equals $1$). This is true even though 
$D$ is Grassmann even. For simplicity, we shall denote $\delta_e$
by $\delta$ from now on.

\subsection{The BRST variation of ${\cal U}$}

We now proceed to compute the BRST variation of ${\cal U}$. As for the
case of ordinary connections, we find the following formula for the variation of $H$ 
under an infinitesimal change of $M$ (see Appendix \ref{varH}):
\be
\label{Hvar}
\delta \Str(H(0))=-\Str(H(0)I_C(\delta M))~~,
\ee
where:
\be
\label{I_C}
I_C(\delta M)=\int_0^l{d\tau U(\tau)^{-1}\delta M(\tau) U(\tau)}~~.
\ee
Here $U(\tau):=U(\tau,0)$. 

Notice that quantity (\ref{M}) can be written as:
\be
\label{Mdecomp}
M={\hat {\cal A}}+\Delta
\ee
where ${\hat {\cal A}}d\tau$ is the matrix of the direct sum connection 
${\cal A}=\left[\ba{cc}{\cal A}^{(+)}&0\\&{\cal A}^{(-)}\ea\right]$ 
on the circle and:
\be
\Delta:=D^2+\frac{1}{2}\rho^i_0\partial_i D~~.
\ee
We have:
\be
{\hat {\cal A}}={\dot \phi}^{\bar i}A_{\bar i}+\frac{1}{2}F_{{\bar
    i}j}\eta^{\bar i}\rho^j_0~~.
\ee
Here $A$ is the direct sum connection on $End(E)$ introduced in (\ref{AD}). 

The BRST variation of $M$ is given by:
\be
\label{eq1}
\delta M=\delta {\hat {\cal A}}+\delta \Delta
\ee
where:
\be
\label{eq2}
\delta {\hat {\cal A}}=A_{\bar i}{\dot \eta}^{\bar i}+\partial_{\bar i}A_{\bar
  j}\eta^{\bar i}{\dot \phi}^{\bar j}+F_{i{\bar j}}{\dot \phi}^i\eta^{\bar
  j}+\frac{1}{2}\partial_{\bar i}F_{{\bar j}k}\eta^{\bar i}\eta^{\bar j}\rho_0^k~~
\ee
and:
\be
\label{eq3}
\delta \Delta={\dot \phi}^i\partial_i D+\eta^{\bar i}\partial_{\bar i}(D^2)+
\frac{1}{2}\eta^{\bar i}\rho_0^j\partial_{\bar i}\partial_j D~~.
\ee
Let us write:
\be
\label{eq4}
A_{\bar i}{\dot \eta}^{\bar i}:=\frac{d}{d\tau}(A_{\bar i}\eta^{\bar
  i})-\partial_{\bar j}A_{\bar  i}\eta^{\bar i}{\dot \phi}^{\bar j}
-\partial_iA_{\bar j}{\dot \phi}^i\eta^{\bar j}
\ee
so that:
\be
\label{eq5}
\delta {\hat {\cal A}}=\frac{d}{d\tau}(A_{\bar i}\eta^{\bar i})+(\partial_{\bar
  i}A_{\bar j}-\partial_{\bar j}A_{\bar i})\eta^{\bar i}{\dot \phi}^{\bar j}+
\frac{1}{2}\partial_{\bar i}F_{{\bar j}k}\eta^{\bar i}\eta^{\bar j}\rho_0^k~~.
\ee
To arrive at this relation, we used $F_{i{\bar j}}=\partial_iA_{{\bar j}}$,
  which holds because $A_i=0$ for all $i$ (remember that $A$ is a $(0,1)$
  connection). 

We next notice that:
\be
\label{eq6}
U^{-1}\frac{d}{d\tau}(A_{\bar i}\eta^{\bar  i})U=\frac{d}{d\tau}
(U^{-1}A_{\bar i}\eta^{\bar  i}U)+U^{-1}[A_{\bar i}\eta^{\bar i}, M]U~~,
\ee
where we used the relations:
\be
\label{Mderivatives}
\frac{d}{d\tau} U=-MU~~,~~\frac{d}{d\tau}U^{-1}=U^{-1}M~~.
\ee
In equation (\ref{eq6}) and below, the symbol $[\cdot,\cdot]$ denotes the usual commutator.
Remembering equation (\ref{Mdecomp}), we find:
\be
\label{eq7}
[A_{\bar i}\eta^{\bar i}, M]=[A_{\bar i}\eta^{\bar i}, {\hat {\cal A}}]+
[A_{\bar i}\eta^{\bar i}, \Delta]~~,
\ee
with:
\be
\label{eq8}
[A_{\bar i}\eta^{\bar i}, {\hat {\cal A}}]=[A_{\bar i}, A_{\bar j}]\eta^{\bar
  i}{\dot \phi}^{\bar j}+\frac{1}{2}[A_{\bar i}, F_{{\bar j}k}]\eta^{\bar
  i}\eta^{\bar j}\rho_0^k~~
\ee
and:
\be
\label{eq9}
[A_{\bar i}\eta^{\bar i}, \Delta]=\eta^{\bar i}[A_{\bar i},D^2]+\frac{1}{2}\eta^{\bar
  i}\rho_0^j[A_{\bar i}, \partial_jD]~~.
\ee
Combining (\ref{eq5}), (\ref{eq6}) and (\ref{eq7}) gives:
\bea
&&U^{-1}\delta {\hat {\cal A}} U=\\
&&\frac{d}{d\tau}(U^{-1}A_{\bar i}\eta^{\bar  i}U)+
U^{-1}\left((\partial_{\bar
  i}A_{\bar j}-\partial_{\bar j}A_{\bar i})\eta^{\bar i}{\dot \phi}^{\bar j}+
\frac{1}{2}\partial_{\bar i}F_{{\bar j}k}\eta^{\bar i}\eta^{\bar
  j}\rho_0^k+[A_{\bar i}\eta^{\bar i},{\hat {\cal A}}]+[A_{\bar i}\eta^{\bar
  i},\Delta]\right)U~~.\nn
\eea
Using (\ref{eq8}) in this expression and combining with (\ref{eq1}) leads to:
\bea
\label{eq10}
\!\!\!\!\!\!\!\!&&\!\!\!\!U^{-1}\delta MU=\frac{d}{d\tau}(U^{-1}A_{\bar i}\eta^{\bar  i}U)+\\
\!\!\!\!\!\!&&\!\!\!\!U^{-1}\left( F_{{\bar i}{\bar j}}\eta^{\bar i}{\dot \phi}^{\bar
  j}-\frac{1}{4}\partial_k F_{{\bar i}{\bar j}} \eta^{\bar i}\eta^{\bar
  j}\rho_0^k+\delta \Delta+[A_{\bar i}\eta^{\bar i}, \Delta]\right)U~~,\nn
\eea
where we used the Bianchi identities for $F$ in order to simplify the second
term within the round brackets. Combining (\ref{eq3}) and
  (\ref{eq9}), we find:
\be
\label{eq11}
\delta \Delta+[A_{\bar i}\eta^{\bar i}, \Delta]={\dot \phi}^i\partial_i
  D+\eta^{\bar i}\nabla_{\bar i}(D^2)+\frac{1}{2}\eta^{\bar
  i}\rho_0^j(\partial_{\bar i}\partial_j D+[A_{\bar i},\partial_jD])
\ee
where $\nabla_{\bar i}(D^2)=\partial_{\bar i}(D^2)+[A_{\bar i},D^2]$ is the 
covariant derivative of $D^2$ with respect to $A$. 

We next want to re-express the term $U^{-1}{\dot \phi}^i\partial_i
  DU$. Noticing that:
\be
{\dot D}={\dot \phi}^i\partial_i  D+{\dot \phi}^{\bar i}\partial_{\bar i}
  D~~,
\ee
we obtain:
\be
\label{eq12}
U^{-1}{\dot \phi}^i\partial_i  DU=U^{-1}{\dot D}U-U^{-1}{\dot \phi}^{\bar i}\partial_{\bar i}
  DU=\frac{d}{d\tau}(U^{-1}DU)+U^{-1}([D,M]-{\dot \phi}^{\bar i}\partial_{\bar i}
  D)U~~,
\ee
where again we used equations (\ref{Mderivatives}). 
The commutator appearing in the expression above is given by:
\be
\label{example2}
[D,M]={\dot \phi}^{\bar i}[D,A_{\bar i}]+\frac{1}{2}[D,F_{{\bar
  i}j}]\eta^{\bar i}\rho_0^j-\frac{1}{2}\rho_0^i\partial_i(D^2)~~.
\ee
The fact that $D$ anticommutes with $\rho_0^i$ is crucial for obtaining the last
  term (see the discussion in the previous subsection). 
Combining with (\ref{eq12}) gives:
\be
\label{eq13}
U^{-1}{\dot \phi}^i\partial_i  DU=\frac{d}{d\tau}(U^{-1}DU)+
U^{-1}\left(-{\dot \phi}^{\bar i}\nabla_{\bar i}D +\frac{1}{2}[D,F_{{\bar
  i}j}]\eta^{\bar i}\rho_0^j-\frac{1}{2}\rho_0^i\partial_i (D^2)\right)U
\ee
We next substitute this into (\ref{eq11}) to obtain:
\bea
\label{eq14}
U^{-1}(\delta \Delta&+&[A_{\bar i}\eta^{\bar i}, \Delta])U\\
&=&\frac{d}{d\tau}(U^{-1}DU)+U^{-1}\left(\eta^{\bar i}\nabla_{\bar i}(D^2)-{\dot
  \phi}^{\bar i}\nabla_{\bar i}D-\frac{1}{2}\rho_0^i\partial_i(D^2)+\frac{1}{2}
\eta^{\bar i}\rho_0^j\partial_j\nabla_{\bar i}D\right)U~~.\nn
\eea
To arrive at this expression, we used $F_{{\bar i}j}=-\partial_jA_{\bar i}$ in 
(\ref{eq13}) and combined the second term within the round brackets 
of that equation with the third term
in (\ref{eq11}) to produce the last term in (\ref{eq14}). 

Finally, we substitute (\ref{eq14}) into (\ref{eq10}):
\bea
\label{eq15}
&&\!\!\!\!\!U^{-1}\delta MU=\frac{d}{d\tau}\left[U^{-1}(D+A_{\bar i}\eta^{\bar
    i})U\right]+\\
&&\!\!\!\!U^{-1}\left( F_{{\bar i}{\bar j}}\eta^{\bar i}{\dot \phi}^{\bar
  j}-\frac{1}{4}\partial_k F_{{\bar i}{\bar j}} \eta^{\bar i}\eta^{\bar
  j}\rho_0^k+\eta^{\bar i}\nabla_{\bar i}(D^2)-{\dot
  \phi}^{\bar i}\nabla_{\bar i}D-\frac{1}{2}\rho_0^i\partial_i(D^2)+\frac{1}{2}
\eta^{\bar i}\rho_0^j\partial_j\nabla_{\bar i}D \right)U~~.~~~~~~~~~~~\nn
\eea
Using (\ref{eq15}) in equation (\ref{I_C}) gives:
\bea
\label{eq_final}
&&\!\!\!\!I_C(\delta M)=R+\\
&&\!\!\int_0^l{d\tau U^{-1}\left( F_{{\bar i}{\bar j}}\eta^{\bar i}{\dot \phi}^{\bar
  j}-\frac{1}{4}\partial_k F_{{\bar i}{\bar j}} \eta^{\bar i}\eta^{\bar
  j}\rho_0^k+\eta^{\bar i}\nabla_{\bar i}(D^2)-{\dot
  \phi}^{\bar i}\nabla_{\bar i}D-\frac{1}{2}\rho_0^i\partial_i(D^2)+\frac{1}{2}
\eta^{\bar i}\rho_0^j\partial_j\nabla_{\bar i}D \right)U}~~,\nn
\eea
where $R=H(0)^{-1}(D+A_{\bar i}\eta^{\bar i})(l)H(0)-(D+A_{\bar i}\eta^{\bar
    i})(0)$ is the contribution of the total derivative term.  
Substituting this into (\ref{Hvar}) leads to relation (\ref{Qhol}) upon
noticing that the contribution $\Str(H(0)R)$ vanishes because $H(0)$ is even
and due to periodicity of $\phi$ and $\eta$ along the boundary:
\be
\Str(H(0)R)=\Str\left[(D+A_{\bar i}\eta^{\bar i})(l)H(0)\right]-
\Str\left[H(0)(D+A_{\bar i}\eta^{\bar i})(0)\right]=0~~.
\ee 
Here $D(l):=D(\phi(l))$ etc.

\section{Conclusions}

Extending a construction due to \cite{Witten_CS}, we wrote down the general boundary
coupling for the B-twisted Landau-Ginzburg model with an arbitrary
non-compact Calabi-Yau target space. The open  string background 
is a $(0,1)$ superconnection living in a complex vector bundle on the
target. We also showed that the equations of motion
for this background (i.e. the BRST invariance requirement for the partition
function on the worldsheet) amount to the condition that the $(0,\leq 2)$ part
of the superconnection's 
curvature equals a constant endomorphism $c$ plus the identity endomorphism 
multiplied by the Landau-Ginzburg potential $W$. This is a natural deformation of the 
target space equations of motion of the B-twisted sigma model, which are recovered in
the limit $W=0$ and $c=0$ 
and require vanishing of 
the $(0,\leq 2)$ part of the curvature. Our results agree with the intuition
that the basic branes of the topological Landau-Ginzburg model are condensates of
elementary D-branes of the B-model.

\acknowledgments{The author thanks W. Lerche, I. Brunner and M. Herbst for
  collaboration in a related project and A. Klemm for support and interest in
  his work. }

\appendix

\section{Coupling of the B-model to $(0,1)$ connections}
\label{Bmodel}

In this appendix we review the coupling of the $B$-twisted sigma model
\cite{Witten_mirror} to ordinary $(0,1)$
connections as introduced in \cite{Witten_CS}. Consider
the usual B-model on a Riemann surface $\Sigma$ and focus on a single circle boundary
component $C$. The boundary coupling of \cite{Witten_CS} has the form:
\be
\label{Z_B}
Z_B=\int{{\cal D}[\phi]{\cal D}[{\tilde F}]{\cal D}[\theta]{\cal D}[\rho]{\cal
    D}[\eta]e^{-{\tilde S}_B} \Tr{H_A}}
\ee
where:
\be
H_A= Pe^{-\oint_{C}{d\tau {\hat {\cal A}}}}
\ee
is the Wilson loop of the `improved connection':
\be
{\hat {\cal A}}=A_{\bar i}{\dot \phi}^{\bar i}+\frac{1}{2}F_{{\bar i}j}\eta^{\bar i}\rho_0^j~~.
\ee
The circle $C$ is the boundary of $\Sigma$ and 
$A$ is a $(0,1)$ connection on a complex vector bundle $E$ over the
target space $X$. For the B-twisted sigma model, the bulk action ${\tilde S}_B$ remains BRST closed
when considered on bordered Riemann surfaces. Thus 
BRST invariance of the partition function (\ref{Z_B}) requires:
\be
\delta \Tr H_A=0~~.
\ee
As sketched in \cite{Witten_CS}, the BRST variation of the Wilson loop factor has the form:
\be
\label{B_deltaH}
\delta \Tr H_A(0) =-\Tr\left[H_A(0) \int_{0}^l{d\tau U^{-1}\left(F_{{\bar i}{\bar j}}\eta^{\bar
      i}{\dot \phi}^{\bar j} -\frac{1}{4}\partial_k F_{{\bar i}{\bar
        j}}\eta^{\bar i}\eta^{\bar j} \rho_0^k\right) U} \right]~~
\ee
where $U(\tau)$ is the parallel transport operator of ${\cal A}={\hat {\cal
      A}}d\tau$ along $C$, starting
      from a distinguished point on the boundary which defines the origin of
      the proper length coordinate $\tau$. Here $l$ is the circumference of
      $\tau$ measured with respect to the metric induced from the interior of
      $\Sigma$. Thus BRST invariance of (\ref{Z_B}) requires that $A$ is an
      integrable connection.

Let us give the proof of equation (\ref{B_deltaH}). 
First notice that ${\cal A}$
is a connection on the complex pulled-back bundle ${\cal E}=\phi_\partial^*(E)$, which
can be trivialized over $C$ (here $\phi_{\partial}$ is the restriction of
$\phi$ to the boundary $\partial \Sigma$). Hence we can view 
${\hat {\cal A}}$ as a matrix-valued function on the circle. 
Thus $U(\tau)$ is uniquely determined by the equation:
\be
\label{hol_eq}
\left(\frac{d}{d\tau}+{\hat {\cal A}}\right)U=0~~
\ee
and the initial condition $U(0)={\rm id}$. 
The holonomy operator at the origin is given by $H_A(0):=U(l)$,
where $l$ is the length of $C$. Varying the origin changes $H_A(0)$ by a
similarity transformation, but does not affect its trace.

Let us consider the change of $H_A(0)$ induced by an arbitrary variation
of ${\cal A}$. Taking the variation of (\ref{hol_eq}) gives:
\be
\left(\frac{d}{d\tau}+{\hat {\cal A}}\right)\delta U=-\delta {\hat {\cal A}} U~~.
\ee
This is equivalent with:
\be
\label{Phi_eq}
\frac{d}{d\tau}\Phi=-U^{-1}\delta {\hat {\cal A}} U
\ee
where we introduced the quantity $\Phi:=U^{-1}\delta U$. The initial condition
$U(0)={\rm id}$ gives $\delta U(0)=0$ and thus $\Phi(0)=0$. With this 
constraint, equation (\ref{Phi_eq}) is solved by:
\be
\Phi(\tau)=-\int_0^{\tau}{ds U(s)^{-1}\delta {\hat {\cal A}}(s)U(s)}
\ee
so that:
\be
\delta U(\tau)=-U(\tau) \int_0^{\tau}{ds U(s)^{-1}\delta {\hat {\cal A}}(s)U(s)}~~.
\ee
This gives:
\be
\label{traced_hol_var}
\Tr \delta H_A(0)=-\Tr\left[H_A(0)\int_0^l{d\tau U(\tau)^{-1}\delta 
{\hat {\cal A}} (\tau) U(\tau)}\right]~~.
\ee
It is easy to check that the right hand side is independent of the choice of
origin for $\tau$. 

To recover (\ref{B_deltaH}), we must apply this formula for the BRST variation of
${\hat {\cal A}}$. The BRST transformations of the B-twisted sigma model are
given by (\ref{BRST}) with $W$ set to zero. We have:
\be
\label{deltaA}
\delta {\hat {\cal A}}=A_{\bar i}{\dot \eta}^{\bar i}+\partial_{\bar i}A_{\bar
  j}\eta^{\bar i}{\dot \phi}^{\bar j}+F_{i{\bar j}}{\dot
  \phi}^i\eta^{\bar j}+\frac{1}{2}\partial_{\bar i} F_{{\bar j}
  k}\eta^{\bar i}\eta^{\bar j} \rho_0^k~~,
\ee
where the dot stands for $\frac{d}{d\tau}$.  

To eliminate the $\tau$-derivative of $\eta$ in the first term, we write:
\be
U^{-1}A_{\bar i}{\dot \eta}^{\bar i}U=U^{-1}\frac{d}{d\tau} (A_{\bar
i}\eta^{\bar i})U-U^{-1}\left[\partial_i A_{\bar j}{\dot \phi}^i\eta^{\bar
  j}+\partial_{\bar j}A_{\bar i}\eta^{\bar i}{\dot \phi}^{\bar j}\right]U
\ee
and:
\be
\label{Eq1}
U^{-1}\frac{d}{d\tau} (A_{\bar i}\eta^{\bar i})U=\frac{d}{d\tau}(U^{-1}A_{\bar
  i}\eta^{\bar i}U)+
U^{-1}[A_{\bar i}\eta^{\bar i}, {\hat {\cal A}}]U~~,
\ee
where we used the relations:
\be
\frac{d}{d\tau}U=-{\hat {\cal A}} U~~,~~\frac{d}{d\tau} U^{-1}=U^{-1}{\hat {\cal A}}~~.
\ee 
The commutator in (\ref{Eq1}) is given by:
\be
[A_{\bar i}\eta^{\bar i}, {\hat {\cal A}}]=[A_{\bar i}, A_{\bar j}]\eta^{\bar
  i}{\dot \phi}^{\bar j}+\frac{1}{2}[A_{\bar i}, F_{{\bar j}k}]\eta^{\bar
  i}\eta^{\bar j}\rho_0^k~~.
\ee

Combining everything, we find:
\be
\label{UdAU}
U^{-1}\delta {\hat {\cal A}} U=\frac{d}{d\tau}(U^{-1}A_{\bar i}\eta^{\bar i}U)+
U^{-1}\left[F_{{\bar i}{\bar j}}\eta^{\bar i}{\dot
    \phi}^{\bar j}-
\frac{1}{4}\partial_k F_{{\bar i}{\bar j}}\eta^{\bar i}\eta^{\bar j}\rho_0^k\right]U~~,
\ee
where 
\be
F_{{\bar i}{\bar j}}=\partial_{\bar i}A_{\bar j}-\partial_{\bar
  j}A_{\bar i}+[A_{\bar i}, A_{\bar j}]
\ee
is the $(0,2)$ part of the curvature of $A$.
To arrive at (\ref{UdAU}), we used the Bianchi identities and the relation $F_{i
    {\bar j}}=\partial_i A_{\bar j}$ (which holds because $A$ is a $(0,1)$
  connection). Using (\ref{UdAU})  into (\ref{traced_hol_var}) leads to 
  (\ref{B_deltaH}) upon noticing that the boundary term induced by the
  total derivative in (\ref{UdAU}) brings vanishing contribution  to
  (\ref{traced_hol_var}) due to the periodicity of $\phi$ and
  $\eta$ along the boundary.

\section{Variation of $H$ with respect to $M$}
\label{varH}

In this appendix we derive relation (\ref{Hvar}), which gives the infinitesimal
change of the `holonomy' operator $H$ under a variation of $M$. 
We shall use the notation $U(\tau):=U(\tau,0)$ as in Section 4. The argument
is very similar to that of Appendix \ref{Bmodel}.

Remember from (\ref{Mderivatives}) that $U(\tau)$ satisfies:
\be
\label{Mhol_eq}
\left(\frac{d}{d\tau}+M\right)U=0~~
\ee
with the initial condition $U(0)={\rm id}$. 

Consider the change of $H(0)$ induced by a variation
of $M$. Taking the variation of (\ref{Mhol_eq}) gives:
\be
\left(\frac{d}{d\tau}+M\right)\delta U=-\delta M U~~.
\ee
This is equivalent with:
\be
\label{MPhi_eq}
\frac{d}{d\tau}\Phi=-U^{-1}\delta M U
\ee
where we introduced $\Phi:=U^{-1}\delta U$. The constraint
$U(0)={\rm id}$ gives $\delta U(0)=0$ and thus $\Phi(0)=0$. With this initial
condition, equation (\ref{Phi_eq}) is solved by:
\be
\Phi(\tau)=-\int_0^{\tau}{ds U(s)^{-1}\delta M(s)U(s)}~~,
\ee
which gives:
\be
\delta U(\tau)=-U(\tau) \int_0^{\tau}{ds U(s)^{-1}\delta M(s)U(s)}~~.
\ee
Applying this for $\tau=l$ and recalling that $U(l)=U(l,0)=H(0)$, we find:
\be
\delta H(0)=-H(0) \int_0^{l}{ds U(s)^{-1}\delta M(s)U(s)}~~.
\ee 
This implies equation (\ref{Hvar}).

\end{document}